\documentclass[pra,aps,floatfix,showpacs,twocolumn]{revtex4-1}

\usepackage{graphicx}
\usepackage{amsmath}
\usepackage{amssymb}
\usepackage{bm}
\usepackage{maybemath}
\usepackage{color}
\usepackage{dcolumn}

\newcolumntype{.}{D{x}{}{6}}

\newcommand{\vp}{\mathrm{vp}}

\newcommand{\meV}{\mathrm{meV}}

\newcommand{\muH}{$\mu$H{}}
\newcommand{\muD}{$\mu$D{}}
\newcommand{\muHeThree}{$\mu$He$^3$}
\newcommand{\muHeFour}{$\mu$He$^4$}

\newcommand{\vecpt}{\vec p^{\,2}}

\newcommand{\dd}{\mathrm{d}}
\newcommand{\ee}{\mathrm{e}}
\newcommand{\ii}{\mathrm{i}}

\begin{document}

%
%
\title{Relativistic Reduced--Mass and Recoil Corrections to Vacuum Polarization\\
in Muonic Hydrogen, Muonic Deuterium and Muonic Helium Ions}

\author{U. D. Jentschura}

\affiliation{Department of Physics,
Missouri University of Science and Technology,
Rolla, Missouri, MO65409, USA}

\begin{abstract}
The reduced-mass dependence of relativistic and radiative
effects in simple muonic bound systems is investigated.
The spin-dependent nuclear recoil correction
of order $(Z\alpha)^4 \, \mu^3/m_N^2$ is evaluated 
for muonic hydrogen and deuterium, and 
muonic helium ions
($\mu$ is the reduced mass and $m_N$ is the nuclear mass).
Relativistic corrections to vacuum polarization of
order $\alpha (Z\alpha)^4 \mu$ are calculated, with a full account 
of the reduced-mass dependence. The results shift theoretical predictions.
The radiative-recoil correction to 
vacuum polarization of order 
$\alpha (Z\alpha)^5 \, \ln^2(Z \alpha)\mu^2/m_N$ 
is obtained in leading logarithmic approximation. 
The results emphasize the need for a unified treatment of relativistic corrections
to vacuum polarization in muonic hydrogen, muonic deuterium and 
muonic helium ions, where the mass ratio of
the orbiting particle to the nuclear mass is larger than the fine-structure
constant.
\end{abstract}


\maketitle


%
%
\section{Introduction}

In muonic hydrogen and muonic deuterium,
the mass ratio $\xi_N = m_\mu/m_N$ 
of the orbiting particle (muon mass $m_\mu$) to the 
mass of the atomic nucleus $m_N$
is  not really small against unity.
It evaluates to 
\begin{subequations}
\begin{align}
\xi_p = &\; \frac{m_\mu}{m_p} = 0.112609\ldots \approx \tfrac{1}{9} \,,
\\[0.77ex]
\xi_d =& \; \frac{m_\mu}{m_d} = 0.0563327\ldots \approx \tfrac{1}{18},
\end{align}
where the latest recommended values of the masses have been 
used~\cite{MoTaNe2008}.
For muonic helium ions, we have
\begin{align}
\label{xiHe3}
\xi_{\rm He^3} =& \; \frac{m_\mu}{m_{\rm He^3}} = 
0.0376223\ldots \approx \tfrac{1}{26} \,,
\\[0.77ex]
\label{xiHe4}
\xi_{\rm He^4} = & \; \frac{m_\mu}{m_{\rm He^4}} = 
0.0283465\ldots \approx \tfrac{1}{35} \,.
\end{align}
\end{subequations}
In all cases, $\xi_N$ is larger than the fine-structure constant
$\alpha \approx 1/137.036$ that governs the relativistic and 
quantum electrodynamic (QED) effects. 
Consequently, the reduced-mass dependence of all quantum 
electrodynamic (QED) effects that influence the spectrum must be
taken into account exactly, i.e., to all orders.
In calculations, one must first
take into account $\xi_N$ (if possible) to all orders, before advancing to the 
next order in the $Z\alpha$-expansion; otherwise the 
higher-order effects in $Z\alpha$ will be shadowed by 
the unknown reduced-mass dependence of lower-order
terms in the $Z\alpha$-expansion.

Hence, particular emphasis has been laid in Ref.~\cite{Pa1996mu} on the correct
treatment of the reduced-mass dependence of all relativistic and QED
corrections.  The statement made in the text preceding Eq.~(17) in
Ref.~\cite{Pa1996mu}, which says ``the external field approximation does
not give an accurate result,'' can hardly be overemphasized. Here,  the
external field approximation refers to the Dirac equation, which is appropriate
for heavy muonic atoms where the parameter $Z\alpha$ (with $Z$ denoting the
nuclear charge number) is much larger than the mass ratio $m_\mu/m_N$, where
$m_N$ is the mass of the heavy nucleus. Even a tiny conceivable error in the
handling of, say, the reduced-mass dependence of the one-loop vacuum
polarization (VP) shift in  muonic hydrogen could drastically influence the
comparison of theory and experiment: the current discrepancy~\cite{PoEtAl2010}
of theory and experiment for the muonic hydrogen Lamb shift amounts to roughly
$0.3 \, \meV$, which is about one part per thousand of the leading
vacuum-polarization contribution and thus smaller than a conceivable additional
reduced-mass correction to the leading VP effect of relative order $\xi_N^3$.

In comparison to previous studies on heavy muonic atoms and ions (excellent
theoretical overviews are provided in Refs.~\cite{BrMo1978,BoRi1982}), the
magnitude of the mass ratio is the main characteristic property of muonic
hydrogen and deuterium.  In this article, we thus revisit the precise treatment
of the vacuum-polarization contribution to the Lamb shift in muonic hydrogen
(\muH) and muonic deuterium (\muD), as well as muonic helium ions (\muHeThree{}
and \muHeFour{}), with full account of the two-body structure of the bound
system.  Starting from the nonrelativistic Hamiltonian (Sec.~\ref{nonrel}), we
proceed to discuss the nuclear-spin dependent terms in the Breit Hamiltonian
(Sec.~\ref{breitbg}), before proceeding to the radiatively corrected Breit
Hamiltonian (Sec.~\ref{radcorr}) and the radiative-recoil correction
(Sec.~\ref{recvp}). Conclusions are drawn in Sec.~\ref{conclu}.

%
%
\section{Nonrelativistic Hamiltonian}
\label{nonrel}

The nonrelativistic
\muH{} Hamiltonian is separable, and
the nonrelativistic (Schr\"{o}dinger) Hamiltonian 
in the center-of-mass system, where the muon and the nuclear particle
carry opposite momenta $\vec p$ and $-\vec p$, respectively, reads
(in natural units, $\hbar = c = \epsilon_0 = 1$),
\begin{equation}
H = \frac{\vecpt}{2 m_\mu} + \frac{\vecpt}{2 m_N}  - \frac{Z\alpha}{r} =
\frac{\vecpt}{2 \mu} - \frac{Z\alpha}{r} \,,
\quad 
\mu = \frac{m_\mu}{1 + \xi_N} \,.
\end{equation}
This equation can be solved exactly in terms of Schr\"{o}dinger
eigenstates. The nonrelativistic spinor wave functions for the 
$2S_{1/2}$ and $2P_{1/2}$ 
states are exact eigenstates of $H$ and read,
explicitly,
\begin{align}
\label{nrwave}
\psi_{2S}(\vec r) =& \; \frac{(Z\alpha \mu)^{3/2}}{2 \sqrt{2}} \, 
(2 - Z\alpha\mu r) \ee^{-\frac12 Z\alpha\mu r} \,
\chi_{-1}^{M}(\hat r) \,,
\\[0.11ex]
\psi_{2P_{1/2}}(\vec r) =& \; \frac{(Z\alpha \mu)^{5/2} \, r}{2 \sqrt{6}} \,  
\ee^{-\frac12 Z\alpha\mu r} \,
\chi_{+1}^{M}(\hat r) \,, 
\end{align}
where $M = \pm \tfrac12$ is the magnetic projection,
$\chi_{\varkappa}^{M}(\hat r)$ is the standard
two-component spin-angular function~\cite{BeSa1957}, and 
$\varkappa = (-1)^{j + \ell + 1/2}$ is the Dirac angular quantum number. 
The reduced-mass dependence
of the wave functions in Eq.~\eqref{nrwave} is exact.

The one-loop vacuum-polarization potential $V_\vp$ can
be expressed in terms of the action of a linear operator $K$ on a 
screened Coulomb potential $v_\vp$ as follows,
\begin{equation}
V_\vp(r) = K[ v_\vp(m_e \, \rho; r) ] \,, \;\;
v_\vp(\lambda; r) = -\frac{Z\alpha}{r} \, \ee^{-\lambda \, r} \,,
\end{equation}
with
\begin{equation}
K[f(\rho)] = \frac{2 \alpha}{3 \pi} \int\limits_2^\infty \dd \rho \;
 \frac{2 + \rho^2 }{\rho^3} \sqrt{1 - \frac{4}{\rho^2}}  \; f(\rho),
\end{equation}
where $m_e$ is the electron mass.  In the following, 
we often use the identification $\lambda = m_e \, \rho$ and 
define the ratio 
\begin{equation}
\beta_N = \frac{m_e}{(Z \alpha \, \mu)} \,,
\end{equation}
which evaluates to
$\beta_p = 0.7373836\ldots$ for \muH{} and
$\beta_d = 0.7000861\ldots$ for \muD{}
(proton and deuteron nuclei, respectively).
For muonic helium ions, the values are
$\beta_{\rm He^3} = 0.3438429\dots$ and $\beta_{\rm He^4} = 0.3407691\dots$.
(We here refrain from assigning a subscript to the 
reduced mass $\mu$, 
even though it of course depends on the nucleus $N$,
because the symbol $\mu_N$ is reserved, canonically,
for the nuclear magnetic moment.)
We then use the exact nonrelativistic unperturbed wave functions
defined in Eq.~\eqref{nrwave} and calculate
the  leading VP energy shifts as
\begin{subequations}
\begin{equation}
\langle 2S_{1/2} | V_\vp | 2S_{1/2} \rangle =
-(Z\alpha)^2 \, \mu \, K\left[ \frac{2 \beta_N^2 \rho^2 + 1}%
{4 \left(\beta_N \, \rho + 1 \right)^4} \right]
\end{equation}
and 
\begin{equation}
\langle 2P_{1/2}  | V_\vp | 2P_{1/2} \rangle = 
-(Z\alpha)^2 \, \mu \, 
K\left[ \frac{1}{4 \left( 1 + \beta_N \, \rho\right)^4} \right]  \,.
\end{equation}
\end{subequations}
A numerical evaluation of these compact expressions
is found to be in agreement with 
the literature (see Refs.~\cite{PoEtAl2010,Pa1996mu,Bo2005mup,Bo2005mud})
and confirms that the reduced-mass dependence of the leading VP effect
is correctly described by Schr\"{o}dinger wave functions scaled
with the reduced mass of the system.
It is even possible~\cite{Pu1957,Ka1998muonic} to carry out the integration 
over the spectral parameter $\rho$ analytically, with the result
\begin{align}
& \langle 2P_{1/2} | V_\vp | 2P_{1/2} \rangle -
\langle 2S_{1/2} | V_\vp | 2S_{1/2} \rangle 
\nonumber\\[2ex]
& = \frac{\alpha}{\pi} \, (Z\alpha)^2 \, \mu \left[
\frac{8 \pi \beta_N^3}{3} 
+ \frac{1 - 26 \beta_N^2 + 352 \beta_N^4 - 768 \beta_N^6}%
{18 \, (1 - 4 \, \beta_N^2)^2} \right.
\nonumber\\
& \; \left. + \frac{4 \beta_N^4 \left( 15 - 80\beta_N^2 + 128\beta_N^4 \right) }%
{3 \, (1 - 4 \, \beta_N^2)^{5/2}} \,
\ln\left( \frac{1- \sqrt{1- 4\beta_N^2}}{2 \beta_N} \right) \right] 
\end{align}
for the Lamb shift difference of the leading VP energy correction.
For reference, the $2P_{1/2}$--$2S_{1/2}$ difference of the 
leading nonrelativistic vacuum polarization 
effect is $205.0073 \, \meV$ for \muH{},
$227.6346\,\meV$ for \muD{},
$1641.885\,\meV$ for \muHeThree{}, and
$1665.772\,\meV$ for \muHeFour{}.
The latter value differs by $0.010 \, \meV$ from the 
value of $1665.782\,\meV$ given in Eq.~(10) of Ref.~\cite{Ma2007};
the difference probably is due to updated physical constants
used in our calculation (see also Ref.~\cite{MoTaNe2008}).

%
%
\section{Breit Hamiltonian and Barker--Glover Terms}
\label{breitbg}

The Breit Equation and the corresponding Hamiltonian
follow from the Bethe--Salpeter equation in the limit on an instantaneous
interaction kernel~\cite{Ma1989} and describe the bound states of general two-body
systems of arbitrary mass ratio $\xi_N$, including higher-order relativistic
corrections~\cite{Ma1991}.
For the $2P_{1/2}$--$2S_{1/2}$ Lamb shift in muonic bound systems, the relevant terms 
in the Breit Hamiltonian read ($\delta_I = 1$ for half-integer 
and $\delta_I = 0$ for integer nuclear spin, see~\cite{PaKa1995})
\begin{align}
\label{deltaH}
\delta H =& \; \sum_{j=1}^{4} \delta H_j \,,
\qquad
\delta H_1 =  - \frac{\vec p^{\,4}}{8 m_\mu^3 } - \frac{\vec p^{\,4}}{8 m_N^3 }  \,,
\nonumber\\
\delta H_2 =& \;  \left( \frac{1}{m_\mu^2} + \frac{\delta_I}{m_N^2} \right) 
\frac{\pi Z \alpha \, \delta^3(r)}{2} \,,
\nonumber\\
\delta H_3 =& \; -\frac{Z\alpha}{2 m_\mu m_N r} \left( \vec p^{\,2} + 
\frac{1}{r^2} r^i r^j p^i p^j \right)  \,,
\nonumber\\
\delta H_4  =& \;  \frac{Z\alpha}{r^3} \,
\left( \frac{1}{4 m_\mu^2} + \frac{1}{2 m_\mu m_N} \right)
\; \vec \sigma \cdot \vec L \,.
\end{align}
where we use the summation convention for the 
superscripts $i$ and $j$ which denote the Cartesian 
components of the position and momentum operators.
Using the relations
\begin{equation}
\vec\nabla^2 \left( \frac{1}{r} \right) = -4\pi \, \delta^3(r)
\end{equation}
and
\begin{equation}
\nabla^i \nabla^j
\left( \frac{x^i \, x^j}{r^3} \right) = +4\pi \, \delta^3(r),
\end{equation}
one may transform $\delta H_3$ to a more symmetric form,
\begin{equation}
\delta H_3 = -\frac{Z\alpha}{2 m_\mu m_N} 
p^i \left( \frac{1}{r} + 
\frac{r^i \, r^j}{r^3} \right) p^j \,.
\end{equation}
After some algebra,
the expectation values of the eigenstates given in Eq.~\eqref{nrwave} 
of the Breit Hamiltonian read  
\begin{subequations}
\label{breit}
\begin{align}
\left< 2S_{1/2} \left| \delta H \right| 2S_{1/2} \right> = & \;
- (Z\alpha)^4 \mu \frac{ 5 + \xi_N (11 + 13\, \xi_N)}{128 \, (1+\xi_N)^2} 
\nonumber\\
& \; + \delta_I \, \frac{(Z\alpha)^4 \; \mu \; \xi_N^2}{16 \, (1 + \xi_N)^2} 
\,,
\\
\left< 2P_{1/2} \left| \delta H \right| 2P_{1/2} \right> = & \;
- (Z\alpha^4) \mu \frac{ 15 + \xi_N (33 + 7\, \xi_N)}{384 \, (1+\xi_N)^2} \,,
\end{align}
\end{subequations}
and the $2P_{1/2}$--$2S_{1/2}$ difference ($2P_{1/2}$ is energetically higher) 
amounts to 
\begin{align}
\label{LmuH}
L(2P_{1/2} \!-\! 2S_{1/2}) =& \; 
\frac{(Z\alpha)^4 \mu \xi_N^2}{48 \, (1 + \xi_N)^2} (4 - 3 \, \delta_I) 
\nonumber\\
=& \; \left\{ \begin{array}{cc} 
\dfrac{(Z\alpha)^4 \mu^3}{48 \, m_N^2}  & \qquad (\delta_I = 1) \,, \\[2.77ex]
\dfrac{(Z\alpha)^4 \mu^3}{12 \, m_N^2}  & \qquad (\delta_I = 0) \,.
\end{array}  \right.
\end{align}
The Barker-Glover~\cite{BaGl1955} correction $L$ 
given in Eq.~\eqref{LmuH} 
evaluates to $0.05747 \, \meV$ for \muH{} and
to $0.12654 \, \meV$ for \muHeThree,
in full agreement with the literature [Eq.~(46) of Ref.~\cite{Pa1996mu}].
Because the zitterbewegung term is absent for the spin-$1$ 
deuteron nucleus~\cite{PaKa1995} and for the spin-zero 
alpha particle, the shift evaluates to $L = 0.06722 \, \meV$ for \muD{} and
to $L = 0.29518\, \meV$ for \muHeFour{} 
[cf.~Eq.~(61) of Ref.~\cite{Ma2007} and 
Eq.~(10) of Ref.~\cite{Bo2011preprint}].
It constitutes a nuclear spin-dependent recoil correction to the 
Lamb shift, which is essential for the correct 
description of the muonic isotope shift.
Equation~\eqref{LmuH} is exact to all orders in $\xi_N$.

\begin{table}[t]
\caption{\label{table1} Detailed breakdown of the 
first-order and second-order individual contributions 
$\delta E^{(1)}_i$ and 
$\delta E^{(2)}_j$ to the relativistic Breit correction 
$\delta E_\vp$ of vacuum polarization for \muH,
\muD{}, and muonic helium ions. All units are meV.}
\begin{tabular}{l....}
\hline
\hline
 & \multicolumn{1}{c}{\muH{}} 
 & \multicolumn{1}{c}{\muD{}} 
 & \multicolumn{1}{c}{\muHeThree} 
 & \multicolumn{1}{c}{\muHeFour} \\
\hline
\multicolumn{5}{c}{\rule[-2mm]{0mm}{6mm} $2P_{1/2}$ [meV]} \\
\hline
$\delta E^{(1)}_1$ & -0.x000558 & -0.x000679 & -0.x020331 & -0.x020970 \\
$\delta E^{(1)}_2$ &  0.x000064 &  0.x000038 &  0.x000467 &  0.x000360 \\
$\delta E^{(1)}_3$ & -0.x000290 & -0.x000181 & -0.x004587 & -0.x003584 \\
$\delta E^{(1)}_4$ & -0.x002026 & -0.x002303 & -0.x085970 & -0.x087587 \\
\hline
\rule[-2mm]{0mm}{6mm}
$\delta E^{(1)}$ & -0.x002811 & -0.x003125 & -0.x110421 & -0.x111781 \\
\hline
$\delta E^{(2)}_1$ & -0.x001124 & -0.x001545 & -0.x099980 & -0.x105132 \\
$\delta E^{(2)}_2$ &  0.x0 &  0.x0 &  0.x0 &  0.x0 \\
$\delta E^{(2)}_3$ & -0.x000269 & -0.x000177 & -0.x008427 & -0.x006624 \\
$\delta E^{(2)}_4$ & -0.x001283 & -0.x001521 & -0.x093497 & -0.x095762 \\
\hline
\rule[-2mm]{0mm}{6mm}
$\delta E^{(2)}$ & -0.x002676 & -0.x003243 & -0.x201904 & -0.x207518 \\
\hline
\rule[-2mm]{0mm}{6mm}
$\delta E_\vp$ & -0.x005486 & -0.x006368 & -0.x312324 & -0.x319300 \\
\hline
\multicolumn{5}{c}{\rule[-2mm]{0mm}{6mm} $2S_{1/2}$ [meV]} \\
\hline
$\delta E^{(1)}_1$ &  0.x029112 &  0.x034636 &  0.x846700 &  0.x872150 \\
$\delta E^{(1)}_2$ & -0.x001928 & -0.x001142 & -0.x014512 & -0.x011243 \\
$\delta E^{(1)}_3$ & -0.x002280 & -0.x001416 & -0.x032734 & -0.x025535 \\
$\delta E^{(1)}_4$ &  0.x0 &  0.x0 &  0.x0 &  0.x0 \\
\hline
\rule[-2mm]{0mm}{6mm}
$\delta E^{(1)}$ &  0.x024904 &  0.x032078 &  0.x799454 &  0.x835372 \\
\hline
$\delta E^{(2)}_1$ & -0.x084996 & -0.x108282 & -2.x875794 & -2.x995690 \\
$\delta E^{(2)}_2$ &  0.x044911 &  0.x053594 &  1.x361115 &  1.x402803 \\
$\delta E^{(2)}_3$ & -0.x009064 & -0.x005539 & -0.x106444 & -0.x082889 \\
$\delta E^{(2)}_4$ &  0.x0 &  0.x0 &  0.x0 &  0.x0 \\
\hline
\rule[-2mm]{0mm}{6mm}
$\delta E^{(2)}$ & -0.x049149 & -0.x060227 & -1.x621122 & -1.x675776 \\
\hline
\rule[-2mm]{0mm}{6mm}
$\delta E_\vp$ & -0.x024245 & -0.x028149 & -0.x821668 & -0.x840404 \\
\hline
\multicolumn{5}{c}{\rule[-2mm]{0mm}{6mm} $2P_{1/2}$--$2S_{1/2}$ [meV] and comparison 
to other work} \\
\hline
\rule[-2mm]{0mm}{6mm}
$\Delta E_\vp$ (this work) &  0.x018759 &  0.x021781 &  0.x509344 &  0.x521104 \\
(Ref.~\cite{VePa2004}) & 0.x0169 & & & \\
(Ref.~\cite{Ma2007}) & & & & -0.x202 \\
(Ref.~\cite{Bo2011preprint})$^a$ & 0.x0169 & 0.x0214 & 0.x495 & 0.x508 \\
\hline
\hline
\end{tabular}
$^a$ A conceptually different approach is used in Ref.~\cite{Bo2011preprint}.
\end{table}

%
%
\section{Radiatively Corrected Breit Hamiltonian}
\label{radcorr}

The massive
Breit interaction uses a strictly static
timelike photon propagator component 
\begin{equation}
G_{00}(\vec{q}) = - \frac{1}{\vec{q}^2 + \lambda^2}
\end{equation}
and spatial components 
\begin{equation}
G_{ij}(\vec{q}) =
- \frac{1}{\vec{q}^2 + \lambda^2}\,
\left[ \delta^{ij} - \frac{q^i \, q^j}{\vec{q}^2 + \lambda^2} \right] \,.
\end{equation}
The spatial components are no longer transverse.
One then follows the standard derivation of the 
Breit interaction 
given in Chap.~83 of Ref.~\cite{BeLiPi1982vol4} but has to 
avoid pitfalls. The derivation 
necessitates the evaluation of Fourier transforms,
the most interesting of which is related to the 
interaction [cf. Eq.~(83.13) of~\cite{BeLiPi1982vol4} and 
Sec.~2 of Ref.~\cite{JeSoIn2002}],
\begin{align}
\label{penultimate}
U(\vec{p}, \vec{q}, \lambda) =& \;
- \frac{4 \pi Z\alpha}{m_\mu \, m_N} \, 
\left[ \frac{\vec{p}^2}{\vec{q}^2 + \lambda^2} 
- \frac{(\vec{p} \cdot \vec{q})^2}{(\vec q^2 + \lambda^2)^2} \right.
\nonumber\\
& \left. + 
\frac{\lambda^2 \vec q^2}{4 (\vec q^2 + \lambda^2)^2}
- \frac{\lambda^2 \vec q \cdot \vec p}{(\vec q^2 + \lambda^2)^2}
\right].
\end{align}
For $\lambda = 0$, the Fourier transform of this expression
with respect to $\vec q$
gives the term $\delta H_3$ in Eq.~\eqref{deltaH}.
For a massive photon, we find
\begin{equation}
\int \frac{{\mathrm d}^3 q}{(2 \pi)^3} \,
U(\vec{p}, \vec{q}, \lambda) 
\ee^{\ii \vec q \cdot \vec r} = \delta v_2(r) + \delta v_3(r),
\end{equation}
where $\delta v_2(r)$ and $\delta v_3(r)$ contribute to
the Breit potential $\delta v_{\vp}$ for massive photon exchange,
\begin{align}
\label{dvvp}
\delta v_{\vp} =& \; K[ \delta v_1 + \delta v_2 + \delta v_3 + \delta v_4 ] \,,
\end{align}
where $\delta v_1(r)$ depends on the nuclear spin,
\begin{align}
\label{deltav1}
\delta v_1 =& \; 
\frac{Z\alpha}{8} \left( \frac{1}{m_\mu^2} + \frac{\delta_I}{m_N^2} \right) \,
\left( 4 \pi \delta^3(r) - \frac{\lambda^2 }{r} \, \ee^{-\lambda r} \right) \,,
\end{align}
and the momentum operators act on the ket state in
\begin{subequations}
\begin{align}
\delta v_2 =& \; -\frac{Z\alpha \lambda^2 \ee^{-\lambda r}}{4 m_\mu m_N r}\,
\left( 1 - \frac{\lambda\,r}{2} + 2 \ii \, \vec r \cdot \vec p \right) \,,
\\[0.11ex]
\delta v_3 =& \; -\frac{Z\alpha \;
\ee^{-\lambda r} }{2 m_\mu m_N r} \;
\left( \vec p^2 + \frac{1 + \lambda\, r}{r^2}  \, r^i r^j p^i p^j \right) \,,
\end{align}
\end{subequations}
whereas the spin-orbit coupling is modified to
\begin{align}
\delta v_4 =& \; Z\alpha \left( \frac{1}{4 m_\mu^2} + \frac{1}{2 m_\mu m_N} \right)
\frac{\ee^{-\lambda r} \, (1 + \lambda r)}{r^3} \; \vec \sigma \cdot \vec L \,.
\end{align}
In the terms $\delta v_2$ and $\delta v_3$, 
all the momentum operators act 
on the ``incoming'' wave function (Dirac ket state),
and the Hamiltonian may be used for the evaluation 
of diagonal matrix elements. For off-diagonal elements, 
it is helpful to symmetrize $\delta v_2$ and $\delta v_3$ with respect to 
outgoing and incoming momenta, effectively replacing 
terms of the form 
$f(\vec r) \, \ii \, \vec r \cdot \vec p$ by the commutator
$\ii \, [ f(\vec r) \vec r, \vec p]$ and 
terms of the form
$f^{ij}(\vec r) \, p^i \, p^j$ by the anticommutator
$\tfrac 12\, \{ f^{ij}(\vec r) , \, p^i \, p^j \}$.
In a second step, using the relation 
$\tfrac12 \, \{ A^2, B \} = A\,B\,A + \tfrac12 \, [A, [A, B]]$,
one obtains an even more symmetric form, with
\begin{subequations}
\begin{align}
\delta w_1 = & \; \delta v_1 \,, \qquad
\delta w_4 = \delta v_4 \,, 
\\[0.11ex]
\delta w_2 =& \; -\frac{Z\alpha \lambda^2 \ee^{-\lambda r}}{4 m_\mu m_N r}\,
\left( 1 - \frac{\lambda\,r}{2} \right) \,,
\\[0.11ex]
\delta w_3 =& \; -\frac{Z\alpha \;
\ee^{-\lambda r} }{4 m_\mu m_N} \;
p^i \; \left( \frac{\delta^{ij}}{r} + \frac{1 + \lambda\, r}{r^3}  \, r^i r^j
\right) \; p^j \,,
\\[0.11ex]
\delta v_{\vp} =& \; K[ \delta w_1 + \delta w_2 + \delta w_3 + \delta w_4 ] \,.
\end{align}
\end{subequations}
The terms $\delta w_2$ and $\delta w_3$ are used in
Eq.~(21) of Ref.~\cite{Pa1996mu}.
The $\alpha (Z\alpha)^4 \, \mu$ relativistic reduced-mass 
correction to vacuum polarization 
then is the sum of four first-order perturbations
$\delta E^{(1)}_i$ and four second-order terms $\delta E^{(2)}_j$,
\begin{subequations}
\begin{align}
\label{deltaEvp}
\delta E_\vp = & \; 
\delta E^{(1)} + \delta E^{(2)} =
\sum_{i=1}^4 \delta E^{(1)}_i +
\sum_{j=1}^4 \delta E^{(2)}_j  \,,
\\[0.11ex]
\label{EE1}
\delta E^{(1)}_i = & \;
K\left[ \left< n \ell_j \left| \delta w_i \right| n \ell_j \right> \right] \,,
\\[0.11ex]
\label{EE2}
\delta E^{(2)}_j = & \;
2 \, K\left[ \left< n \ell_j \left| 
\delta H_j 
\right| \delta \psi_{n \ell_j} \right> \right] \,,
\end{align}
\end{subequations}
where $| \delta \psi_{n \ell_j}\rangle$
is the wave function correction due to VP,
\begin{equation}
\label{wfcorr}
| \delta \psi_{n \ell_j} \rangle =
\left( \frac{1}{E_{n\ell} - H} \right)' \, v_\vp \, | n \ell_j \rangle \,.
\end{equation}
Using a generalization of techniques outlined in Ref.~\cite{Je2009g}, 
the perturbation $\delta \psi_{n \ell_j}$ can be evaluated
analytically. The detailed expressions for the 
reduced Green functions (indicated by a prime) of the 
$2S_{1/2}$ and $2P_{1/2}$ states
have been given in Eqs.~(23) and (24) of Ref.~\cite{Pa1996mu}.
All individual contributions are listed in Table~\ref{table1},
in order to facilitate a numerical comparison with independent 
calculations. For \muH{}, we obtain a result of
$\Delta E_{\rm vp} = 
\delta E_\vp(2P_{1/2}) - \delta E_\vp(2S_{1/2}) = 0.018759 \,\meV$.
This result is not in perfect agreement with 
published values~\cite{Pa1996mu,VePa2004,Bo2011preprint}.
For comparison, the result indicated in Eq.~(25) of Ref.~\cite{Pa1996mu}
reads $0.059\,\meV$; and in Eq.~(25) of Ref.~\cite{VePa2004}
a result of $0.0169 \, \meV$ has been indicated.
In Table~1 on page~8 of Ref.~\cite{Bo2011preprint},
a numerically equivalent result of $0.0169\,\meV$ is given.
(We note that Ref.~\cite{Bo2011preprint} contains many unnumbered
tables; the referenced table is numbered).
The matrix elements of the relativistic recoil 
operator given in Eq.~(7) of Ref.~\cite{Bo2011preprint} are
evaluated using unperturbed wave functions.
All values given in Table~\ref{table1} are nonperturbative in 
the mass ratio and take the wave function correction into account.
A precise comparison of individual contributions to 
the approach of Ref.~\cite{Bo2011preprint} is not possible at present.
As evident from Table~\ref{table1}, there are quite significant 
differences with published values for \muHeFour: e.g., the entries in 
Eqs.~(26)--(29) and Eq.~(41) of Ref.~\cite{Ma2007} add up to a
correction of $-0.202 \, \meV$ for the $2P_{1/2}$--$2S_{1/2}$ Lamb shift in 
\muHeFour, whereas we obtain $+0.521 \, \meV$.

A very important question concerns the verifiability 
of the results. In self-energy calculations~\cite{Pa1993},
a cross-check of the calculation consists
in the cancellation of an overlapping parameter that separates 
different momentum and energy regions of the physical process.
For VP effects in 
muonic systems, no such checks are immediately available.
Here, we note that the entries for the first-order matrix 
elements in Table~\ref{table1} for $\mu$He$^4$ are in full 
agreement with the results given in Eqs.~(26)--(29) of Ref.~\cite{Ma2007}.
For the matrix elements 
needed for $\delta E^{(1)}$, the limit as $\lambda \to 0$ of the matrix elements
$\left< n \ell_j \left| \delta w_i \right| n \ell_j \right>$
can be verified independently, and the calculation 
can otherwise be performed analytically, with ease.
For the  matrix elements needed in the evaluation 
of the second-order effects $\delta E^{(2)}$, 
we can verify the first few terms in the 
asymptotic limit as $\lambda \to 0$, using the relation 
\begin{align}
& 2 \, \left< n \ell_j \left| 
\delta H \left( \frac{1}{E - H} \right)' v_\vp \right| 
n \ell_j  \right> 
\\[0.11ex]
& = 2 \, \left< n \ell_j \left| 
\delta H \left( \frac{1}{E - H} \right)' (Z\alpha) 
\frac{\partial}{\partial (Z\alpha)} \right| 
n \ell_j  \right> 
\nonumber\\[0.11ex]
& \quad - \left< n \ell_j \left| 
\delta H \left( \frac{1}{E - H} \right)' 
Z\alpha \, r \,\right|  n \ell_j  \right>  \, \lambda^2 + {\mathcal O}(\lambda^3) \,.
\nonumber
\end{align}
In deriving this relation, the Hellmann--Feynman 
theorem is useful for the zeroth-order term in $\lambda$.
The wave function perturbation in the term of 
order $\lambda^2$ can be evaluated analytically.

\begin{figure}[t!]
\centerline{\includegraphics[width=1.0\linewidth]{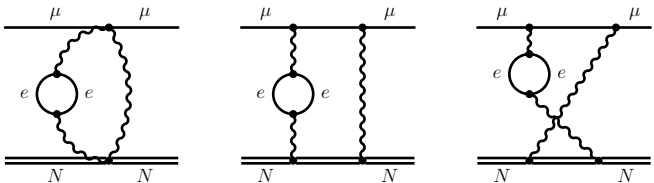}},
\caption{\label{fig1} Feynman diagrams for the 
radiative-recoil correction in two-body muonic bound 
systems. The three given diagrams correspond to the 
vacuum-polarization insertion in the seagull and two-photon 
exchange and lead to the leading double logarithm 
given in Eq.~\eqref{radrec}.}
\end{figure}

%
%
\section{Recoil Corrections to Vacuum Polarization}
\label{recvp}

Beyond 
the radiative modifications of the static Breit 
Hamiltonian, the recoil correction to vacuum polarization can be obtained 
by the insertion of vacuum polarization loops 
into the Salpeter recoil correction~\cite{Sa1952,Er1977,Pa1998}.
The recoil correction is the sum of four terms~\cite{Pa1998};
two of these (low- and middle-energy part) 
describe the frequency-dependent 
part of the Breit interaction, beyond the 
static Breit Hamiltonian given in Eq.~\eqref{deltaH},
and two further terms (seagull and high-energy part) 
correspond to two-photon exchange.

The seagull term corresponding 
to Fig.~\ref{fig1} (left), with a vacuum polarization
insertion in the exchange photon, leads to the integral
\begin{align}
\label{aficion}
& \delta E_S = -\frac{e^4}{2 m_\mu m_N}
K \left[ \int 
\frac{\dd^3 k_1}{(2 \pi)^3} 
\frac{\dd^3 k_2}{(2 \pi)^3} 
\frac{1}{\omega_1 k_2} 
\frac{1}{\omega_1 + k_2} 
\right.
\\[0.11ex]
& \; \left.
\left( \delta^{ij} - \frac{k^i_1 \, k^j_1}{\omega_1^2} \right) \,
\left( \delta^{ij} - \frac{k^i_2 \, k^j_2}{k_2^2} \right)
\right] \, \left< n\ell_j |
\ee^{\ii \, \left( \vec k_1 + \vec k_2 \right) \cdot \vec r} 
| n \ell_j \right> \,,
\nonumber
\end{align}
where $\omega_1 = \sqrt{\vec k_1^2 + \lambda^2}$
is the frequency of the massive photon in the 
vacuum-polarization loop.
An ultraviolet cutoff $\Lambda$ is introduced 
via multiplication of the integrand by 
a multiplicative regularization factor 
$\frac{\Lambda^2}{\vec k_1^2 + \Lambda^2} \,
\frac{\Lambda^2}{\vec k_2^2 + \Lambda^2}$.
The auxiliary parameter $\Lambda$ cancels when the 
high-energy part from two-photon exchange
[see Figs.~\ref{fig1}, middle, and~Fig.~\ref{fig1}, right]
is added to the result (see also Ref.~\cite{Pa1998}).
From the integral~\eqref{aficion}, 
we extract a leading double logarithmic correction,
\begin{equation}
\label{radrec}
\delta E_S =
-\frac{4 \alpha (Z\alpha)^5 \mu^3 \, \delta_{\ell 0}}{3 \pi^2 m_\mu m_N \, n^3} 
\ln^2\left({4 Z\alpha \beta_N^2}\right) \,,
\end{equation}
which is nonvanishing only for $S$ states ($\ell = 0$). 
This correction evaluates to $0.0003 \, \meV$ for the 
$2P_{1/2}$--$2S_{1/2}$ Lamb shift in 
\muH{}, $0.0002 \, \meV$ for \muD{}, $0.0072 \, \meV$ for \muHeThree, 
and $0.0056 \, \meV$ for \muHeFour.
Because subleading logarithmic terms, 
and nonlogarithmic terms are missing, the theoretical uncertainty of the 
results in Eq.~\eqref{radrec} should be taken as 100\,\%
of the leading logarithmic correction calculated here.

\section{Conclusions}
\label{conclu}

Our theoretical investigations are motivated by the
necessity to shed light on the recently observed discrepancy of theory and
experiment in \muH{} (see Ref.~\cite{PoEtAl2010}.  By an explicit evaluation of
the matrix elements of the two-body Breit Hamiltonian, we obtain the
nuclear-spin dependent recoil contributions to the Lamb shift in \muH{} and
\muD{} given in Eq.~\eqref{LmuH}, and confirm that the results are exact in the
mass ratio, so that the existence of further recoil corrections~\cite{BaGl1955}
can be ruled out at order $(Z\alpha)^4$. 
The calculation of the relativistic reduced-mass
corrections to vacuum polarization using the massive Breit Hamiltonian is shown
to involve a nontrivial nuclear-spin dependent term [see Eq.~\eqref{deltav1}]. 
Our detailed numerical
investigation (see Table~\ref{table1}) slightly decreases the observed
experimental-theoretical discrepancy~\cite{PoEtAl2010} (in contrast to a recent
investigation~\cite{CaThRaMi2011}, where the authors obtain an increase of the
discrepancy, based on a treatment which is perturbative in the mass ratio).

A detailed breakdown of the relativistic corrections to vacuum polarization,
including the reduced-mass corrections is given in Table~\ref{table1} for
muonic hydrogen, muonic deuterium and muonic helium ions.  For muonic 
hydrogen, the sum of the entries in rows~3~and~19 of the theory
in the supplemental material of 
Ref.~\cite{PoEtAl2010}, minus the entry in row~1 of the same supplemental material,
amounts to  $(205.0282-0.0041 - 205.0074) \, \meV = 0.0167 \, \meV$; this is
close to the result indicated in Table~1 of Ref.~\cite{Bo2011preprint}, which is
$0.0169 \, \meV$.  In Refs.~~\cite{Bo2005mup,PoEtAl2010,Bo2011preprint}, the
second entry in the mentioned combination ($-0.0041 \, \meV$) has been referred
to as a ``recoil correction to vacuum polarization'', whereas we
here refer to the effect as a relativistic correction
to vacuum polarization with a proper account of the reduced-mass dependence.
Our approach is nonperturbative in the mass ratio $\xi_N$
and isolates the terms of order $\alpha \, (Z\alpha)^4$,
while treating the two-body aspects of the problem to all orders.

Our calculations lead to significant shifts of theoretical predictions for
\muHeFour{} with respect to published values (experiments
are planned for the near future). 
Specifically, for \muHeFour, our nuclear-spin dependent Barker-Glover 
type correction $L$ of $0.295 \, \meV$ differs from the 
value of $0.074\,\meV$ given in Refs.~\cite{Ma2007,Bo2011preprint} 
by $+0.221 \, \meV$.
For \muHeFour{}, our result for the relativistic correction
to vacuum polarization, with a full account of the 
reduced-mass dependence, reads as
$0.521 \, \meV$ for the $2P$--$2S$ difference,
to be compared with a value of $-0.202 \, \meV$ 
given in Ref.~\cite{Ma2007}.
This leads to a total upward shift of theoretical predictions for the
$2P_{1/2}$--$2S_{1/2}$ Lamb shift in \muHeFour{} by 
$[0.521 - (-0.202 ) + 0.221 - 0.010] \meV = +0.934 \, \meV$ relative
to Ref.~\cite{Ma2007} (where we add the small correction of the 
reference value of the leading VP correction)
and by $(+0.221 + 0.013)\, \meV = 0.234 \, \meV$ relative
to Ref.~\cite{Bo2011preprint} (where we add the difference in the relativistic
correction to vacuum polarization from Table~\ref{table1}).

The radiative-recoil correction obtained in
Eq.~\eqref{radrec} is numerically small; however, this two-loop bound-state
correction has traditionally been one of the most elusive effects in
bound-state quantum electrodynamics for two-body systems. Its calculation in
leading logarithmic approximation helps to determine the overall uncertainty of
theoretical predictions with regard to the conceptually involved higher-order
recoil corrections to VP, given by the two-body nature of the bound system.

\section*{Acknowledgments}

Support by NSF and NIST (Precision Measurement Grant),
and helpful conversations with K.~Pachucki are 
gratefully acknowledged.
The author thanks B.~J.~Wundt for carefully reading the 
manuscript.

\end{document}